\documentclass[]{article}

\usepackage[affil-it]{authblk}
\usepackage{a4wide}
\usepackage{amsmath}
\usepackage{multirow}
\usepackage{booktabs}
\usepackage[ruled,vlined]{algorithm2e}
\usepackage{array}
\usepackage{color}
\usepackage[normalem]{ulem}
\usepackage{hyperref}
\usepackage{amssymb}
\usepackage{graphicx}
\usepackage{amsfonts}
\usepackage[english]{babel}
\usepackage[sort&compress,square,comma,authoryear]{natbib}

\begin{document}
\bibliographystyle{apalike}

\title{Searching for pulsars in extreme orbits - GPU acceleration of the Fourier domain 'jerk' search}

\author[1]{Karel Ad\'{a}mek}
\author[1]{Jan Novotn{\'y}}
\author[2]{Sofia~Dimoudi}
\author[1]{Wesley~Armour \thanks{E-mail address: \texttt{wes.armour@oerc.ox.ac.uk}} }
\affil[1]{Oxford e-Research Centre, Department of Engineering Sciences, University of Oxford, 7 Keble road, OX1 3QG, Oxford, United Kingdom}
%\affil[2]{}
\affil[2]{Centre for Advanced Instrumentation, Durham University, South Road, DH1 3LE, Durham, UK;}

\maketitle

%\paperauthor{Karel~Ad\'amek}{}{0000-0003-2797-0595}{University of Oxford}{Oxford e-Research Centre}{Oxford}{Oxfordshire}{OX1 3QG}{UK}
%\paperauthor{Jan~Novotn\'y}{}{0000-0002-9667-635X}{University of Oxford}{Oxford e-Research Centre}{Oxford}{Oxfordshire}{OX1 3QG}{UK}
%\paperauthor{Sofia~Dimoudi}{}{0000-0002-0967-1332}{Durham University}{Centre for Advanced Instrumentation}{Durham}{County Durham}{DH1 3LE}{UK}
%\paperauthor{Wes~Armour}{}{0000-0003-1756-3064}{University of Oxford}{Oxford e-Research Centre}{Oxford}{Oxfordshire}{OX1 3QG}{UK}
% remove/add as you need

Binary pulsars are an important target for radio surveys because they present a natural laboratory for a wide range of astrophysics for example testing general relativity, including detection of gravitational waves. The orbital motion of a pulsar which is locked in a binary system causes a frequency shift (a Doppler shift) in their normally very periodic pulse emissions. These shifts cause a reduction in the sensitivity of traditional periodicity searches. To correct this smearing \citet{2001PhDT.......123R, 2002AJ....124.1788R} developed the Fourier domain acceleration search (FDAS) which uses a matched filtering technique. This method is however limited to a constant pulsar acceleration. Therefore, \citet{2018ApJ...863L..13A} broadened the Fourier domain acceleration search to account also for a linear change in the acceleration by implementing the Fourier domain ''jerk'' search into the PRESTO software package. This extension increases the number of matched filters used significantly. We have implemented the Fourier domain ''jerk'' search (JERK) on GPUs using CUDA. We have achieved 90x performance increase when compared to the parallel implementation of JERK in PRESTO. This work is part of the AstroAccelerate project \citet{AstroAccelerate_2019_2556573}, a many-core accelerated time-domain signal processing library for radio astronomy.
\begin{abstract}
\end{abstract}

\section{Introduction}
Binary pulsars represent examples of some of the most extreme physics in our Universe. These binary systems are natural laboratories for tests of a wide range of astrophysics, from tests of general relativity, including gravitational waves, to tests of equations of state of the matter from which pulsars are made of. However, detection of these pulsars proves to be hard and computationally expensive. The orbital motion of a pulsar which is locked in a binary system causes a frequency shift (a Doppler shift) in the pulsar's normally very periodic pulse emissions. This shift results in a reduction in the sensitivity of traditional periodicity searchers which use the Fourier transform to pick up periodic components in the signal. When the Doppler shifted signal is Fourier transformed the power of the pulsar is smeared into surrounding frequency bins which makes subsequent detection much harder. 

To correct for this smearing, \citet{2001PhDT.......123R, 2002AJ....124.1788R} developed the Fourier domain acceleration search (FDAS) which reduces the smearing by applying a special set of matched filters. However, this approach is limited to constant accelerations, therefore \citet{2018ApJ...863L..13A} expanded this method to account for a linearly changing acceleration, or a constant orbital jerk, of the pulsar. The jerk search increases the number of matched filters used significantly when compared to FDAS. This was implemented into a de-facto software package for processing time-domain radio astronomy data called PRESTO \citep{2011ascl.soft07017R}.

We have implemented a GPU version of the `jerk' search for NVIDIA GPUs using CUDA. This work is based on our previous work on FDAS by \citet{2018ApJS..239...28D} and \citet{2017arXiv171110855A}, where we have used a custom implementation of the FFT algorithm which uses GPU shared memory present on NVIDIA GPUs. This has been is incorporated into AstroAccelerate which also contains other commonly used signal processing algorithms such as de-dispersion \citep{2012ASPC..461...33A} and single pulse search \citep{2019arXiv191008324A}.

\begin{figure}
%\plotone[width=0.70\textwidth]{P10-50_f1}
\centering
\includegraphics[width=0.7\textwidth]{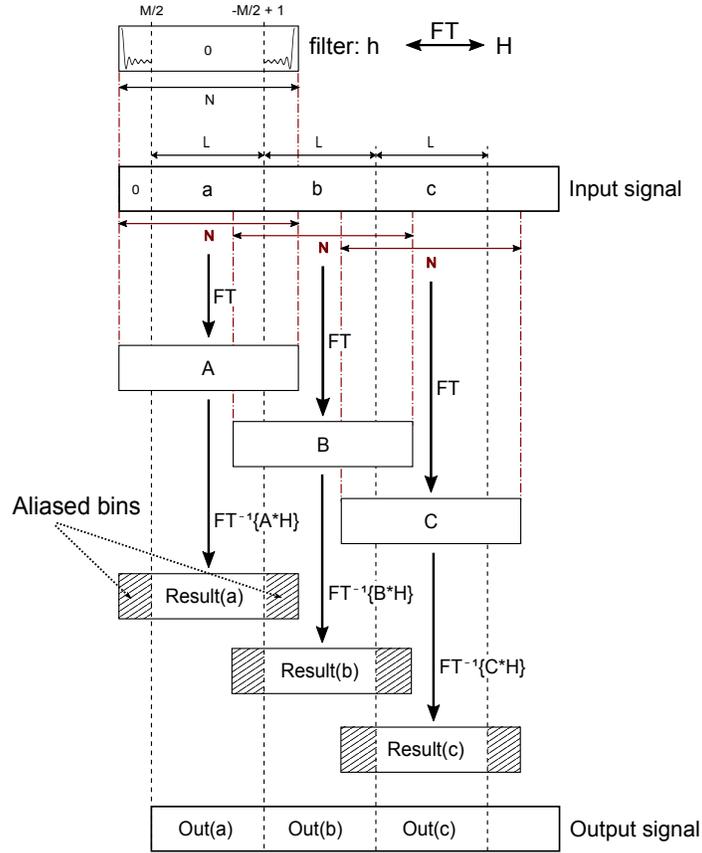}
\caption{Overlap-and-save method: The input signal is separated into overlapping segments (A,B,C,...) which are then processed independently ($FT$ is Fourier transformation). \label{fig:overlapandsave}}
\end{figure}

\section{Implementation}
Our implementation of the `jerk' search expands on our implementation of the FDAS technique, where we have used the overlap-and-save method (shown in Figure \ref{fig:overlapandsave}) to calculate one-dimensional convolutions which uses a shared memory (fastest memory available on NVIDIA GPUs) implementation of the FFT algorithm. This allows us to bypass accesses to the device memory (slowest memory present on the GPU), achieving significant improvements in performance when calculating convolutions. The achieved performance increase depends on the convolution filter length. For filter lengths used in the jerk search, the convolution calculation can be up to 3$\times$ faster than a standard convolution implementation using the NVIDIA cuFFT library for the FFT algorithm. This technique is described in further detail in  \citet{2019arXiv191008324A}. 

\begin{figure}
%\plotone[width=0.75\textwidth]{P10-50_f2}
\centering
\includegraphics[width=0.7\textwidth]{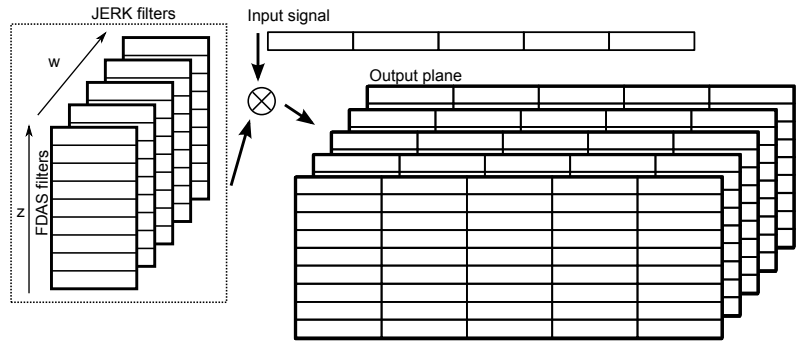}
\caption{Schematic view of the jerk search showing that the extra dimension associated with the linear change in acceleration significantly increases the number of matched filters used. \label{fig:jerksearch}}
\end{figure}

In the 'jerk' search technique, each filter fits a specific Fourier response for a given first frequency derivative $z=\dot{f}T^2$ and second frequency derivative $w=\dot{z}=\ddot{f}T^3$. The number of filters used in the 'jerk' search is greatly increased when compared to the standard Fourier domain acceleration search. In effect jerk searches can be $40\times-80\times$ larger than FDAS searches, depending on sensitivity. This could mean processing as much as 400 GB of data for every 10 minute observation. The jerk search is depicted in Figure \ref{fig:jerksearch} which demonstrates the jerk search size compared to FDAS. Where FDAS requires only one set of filters, the jerk search requires the same number of filters for for every value of linear change in acceleration searched.

\section{Results}
The results presented here are for the matched filtering part of the 'jerk' search technique with candidate selection. The candidate selection involves calculation of the mean and standard deviation and a peak-finding algorithm which selects only peaks which are above some user specified threshold. The execution time of the harmonic sum, which improves results of the Fourier search techniques is not included in our results. The harmonic sum is at least as computationally expensive as the matched filtering part of the jerk search algorithm. 

\begin{figure}
\begin{minipage}{0.49\textwidth}
\centering
\includegraphics[width=0.7\textwidth]{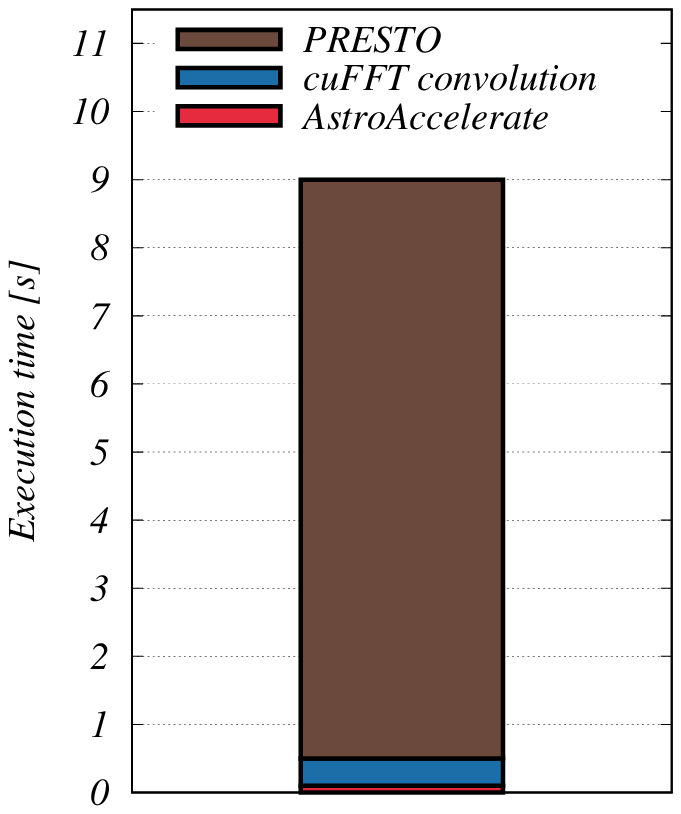}
\end{minipage}
\hfill
\begin{minipage}{0.49\textwidth}
\centering
\includegraphics[width=0.7\textwidth]{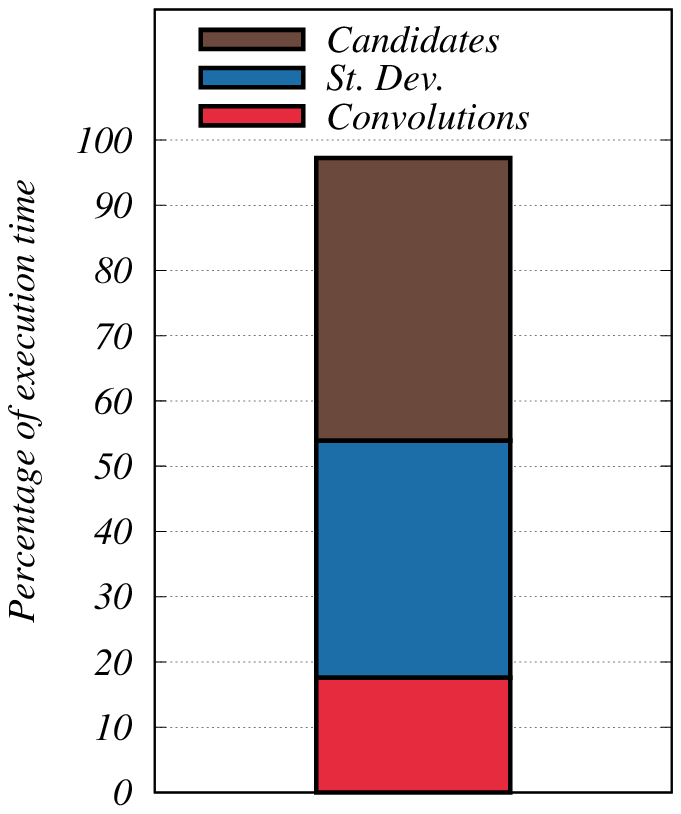}
\end{minipage}
\caption{Comparison of the execution time between AstroAccelerate GPU implementation of the jerk search and PRESTO (left) and breakdown of the execution time of individual algorithmic components (right).  \label{fig:executiontime}}
\end{figure}

For comparison of execution time we have used the NVIDIA Titan V GPU and Intel CPU i7-4770K 3.5GHz. For the CPU version of the 'jerk' search we have used the implementation in PRESTO. The CPU version of the 'jerk' search was run with four threads which was the fastest configuration we have found. A comparison of the execution time of PRESTO with AstroAccelerate is shown in Figure \ref{fig:executiontime} (left). We have also included the execution time of the 'jerk' search which uses the calculation of the convolution based on the NVIDIA cuFFT library. The breakdown of the 'jerk' search execution time to individual components is shown in Figure \ref{fig:executiontime} (right). This shows that the calculation of mean and standard deviation along with candidate selection dominate our jerk search implementation.

Compared to the parallel version of PRESTO our GPU implementation on the NVIDIA Titan V GPU is almost 90$\times$ faster. We see that the calculation of the convolution is only 20\% of the total execution time. Thus, future work will focus on the performance of candidate selection and statistics calculations. 

\section{Conclusion}
We have implemented the jerk search on NVIDIA GPUs using CUDA. Our implementation is almost 90$\times$ faster than the CPU version of the jerk search implemented in PRESTO when tested on a typical desktop CPU. This work is part of the AstroAccelerate package \citep{AstroAccelerate_2019_2556573}.

\bibliography{P10-50}

% if we have space left, we might add a conference photograph here. Leave commented for now.
% \bookpartphoto[width=1.0\textwidth]{foobar.eps}{FooBar Photo (Photo: Any Photographer)}

\end{document}